# Deep learning-enhanced chemiluminescence vertical flow assay for high-sensitivity cardiac troponin I testing

Gyeo-Re Han, [1,†] Artem Goncharov, [1,†] Merve Eryilmaz, [1,2] Shun Ye, [2] Hyou-Arm Joung, [1] Rajesh Ghosh, [2] Emily Ngo, [3] Aoi Tomoeda, [4] Yena Lee, [5] Kevin Ngo, [2] Elizabeth Melton, [6] Omai B. Garner, [7] Dino Di Carlo, [2,8] and Aydogan Ozcan[1,2,8,9]*

[1]Electrical & Computer Engineering Department, [2]Bioengineering Department, [3]Department of Psychology, [4]Chemical and Biomolecular Engineering Department, [5]Mechanical and Aerospace Engineering Department, University of California, Los Angeles, CA 90095 USA, [6]Biomedical Engineering Department, University of California, Davis, CA 95616 USA, [7]Department of Pathology and Laboratory Medicine, [8]California NanoSystems Institute (CNSI), [9]Department of Surgery, University of California, Los Angeles, CA 90095 USA.

[†]These authors contributed equally.

*Corresponding Authors: ozcan@ucla.edu ; dicarlo@ucla.edu

## Abstract

Democratizing biomarker testing at the point-of-care requires innovations that match laboratory-grade sensitivity and precision in an accessible format. Here, we demonstrate high-sensitivity detection of cardiac troponin I (cTnI) through innovations in chemiluminescence-based sensing, imaging, and deep learning-driven analysis. This chemiluminescence vertical flow assay (CL-VFA) enables rapid, low-cost, and precise quantification of cTnI, a key cardiac protein for assessing heart muscle damage and myocardial infarction. The CL-VFA integrates a user-friendly chemiluminescent paper-based sensor, a polymerized enzyme-based conjugate, a portable high-performance CL reader, and a neural network-based cTnI concentration inference algorithm. The CL-VFA measures cTnI over a broad dynamic range covering six orders of magnitude and operates with 50 µL of serum per test, delivering results in 25 min. This system achieves a detection limit of 0.16 pg/mL with an average coefficient of variation under 15%, surpassing traditional benchtop analyzers in sensitivity by an order of magnitude. In blinded validation, the computational CL-VFA accurately measured cTnI concentrations in patient samples, demonstrating a robust correlation against a clinical-grade FDA-cleared analyzer. These results highlight the potential of CL-VFA as a robust diagnostic tool for accessible, rapid cardiac biomarker testing that meets the needs of diverse healthcare settings, from emergency care to underserved regions.

# 1. Introduction

Cardiovascular diseases (CVDs) remain a significant global health issue, responsible for around 19.1 million annual deaths, which accounts for 32% of all global deaths.[1] This significant mortality, coupled with a considerable socioeconomic burden, resulted in CVD-related healthcare costs exceeding $250 billion annually in the United States alone.[2, 3] Notably, CVDs disproportionately impact underserved populations, particularly in low-resource settings where quality preventive and diagnostic services are less accessible.[4, 5] Acute myocardial infarction (AMI), or heart attack, is a serious form of CVD and a leading cause of emergency visits and death, highlighting the need for prompt and accurate diagnostic strategies due to its time-critical nature.[6-8] Diagnosing AMI relies heavily on detecting elevated cardiac troponin I (cTnI) levels, a highly specific and sensitive biomarker of myocardial injury.[9] Blood cTnI levels are often used in conjunction with other diagnostic methods, including patient-reported symptoms, electrocardiograms, and imaging studies, as part of a comprehensive diagnostic framework.[10] Clinical guidelines emphasize the significance of elevated cTnI concentrations beyond established cut-off levels (approximately 10–40 pg/mL) as a necessary criterion in diagnosing AMI.[10, 11] In clinical practice, AMI diagnosis is confirmed when an elevated cTnI level is identified alongside at least one abnormal finding among other supportive diagnostic methods.[10, 11] This approach emphasizes the critical role of accurate and sensitive cTnI quantification as a prioritized factor in guiding AMI diagnosis and management.

Given the critical role of cTnI measurement, numerous cTnI detection methods have been developed, ranging from sophisticated laboratory-based assays to portable point-of-care testing (POCT) platforms. Laboratory-based high-sensitivity cTnI (hs-cTnI) assays have set the gold standard in clinical diagnostics (see **Figure 1a**), as they enable the detection of cTnI at trace levels (down to a few pg/mL) in the bloodstream, facilitating early identification of AMI and accurate patient risk stratification.[12, 13] However, these hs-cTnI assays are mainly confined to centralized hospital laboratories where specialized equipment, skilled personnel, and dedicated bench space are available.[14] The high maintenance and operating costs of these systems can be particularly burdensome for healthcare facilities with limited financial resources.[15] Further, reliance on trained personnel restricts their use in resource-limited settings. Additionally, longer turnaround times associated with the multistep administrative procedures of central laboratory testing can hinder prompt decision-making in critical patient management.[16, 17] These limitations often make hs-cTnI testing inaccessible to patients in underserved areas, where the prevalence of CVDs is higher,[4, 18] or in high-demand emergency settings, where prompt diagnostic decisions are vital.[19] Consequently, the current diagnostic paradigm in AMI care, dominated by laboratory-based testing, faces inherent limitations related to processing time, expense, and accessibility.

Driven by the systemic limitations of laboratory-based cTnI testing, there has been a shift toward patient-centered POCT systems in AMI diagnostics (see **Figure 1b**).[17, 20, 21] POCT enables diagnostic evaluation without requiring extensive infrastructure or specialized equipment, making it valuable in rural, low-resource settings, and in emergency environments, where rapid diagnosis is crucial.[5, 16, 17, 21] There are currently several cTnI-POCT products on the market utilizing technologies such as microfluidics,[22-26] lateral flow tests,[27] and automated immunoassays.[28, 29] Some of the recent advancements in cTnI biosensing also include paper-based tests,[30-40] digital microfluidics,[41] microparticle assays,[42, 43] electrochemical methods,[44-46] and array-based sensing.[47, 48] Although these technologies expand the capabilities of cTnI-POCT, achieving the high sensitivity needed for hs-cTnI detection remains challenging for most POCT platforms.[10, 21] The exceptionally low cTnI cut-off levels required to differentiate AMI from non-AMI cases are difficult for most POCT systems to meet, particularly those designed with low-cost sensors. Despite advances in miniaturization, many POCT systems still retain benchtop-sized footprints and remain costly, limiting their use in smaller, resource-constrained clinical environments.[22, 24, 28, 29] Thus, developing a POCT platform that combines affordability, true portability, and high sensitivity remains a critical goal, as overcoming these barriers could greatly improve diagnostic access and support the shift toward more patient-centered care.

Although current POCT systems have limitations in achieving the necessary sensitivity required for clinical use, chemiluminescence (CL)-based detection has proven effective for highly sensitive biomarker assays. Among the 14 FDA-, CE-, or equivalent regulatory-cleared commercial hs-cTnI benchtop analyzers devised for clinical laboratory use, 12 of them utilize CL-based technologies (see **Table S1**).[49, 50] CL has a distinctive advantage in generating strong, high-contrast signals without the need for an external light source, unlike fluorescence-based methods that require complex excitation optics.[51] Its high signal-to-noise (S/N) ratio, achieved by low background noise, significantly enhances detection sensitivity, while the broad calibration range of CL enables the accurate measurement of cTnI across clinically relevant concentrations—from trace levels (a few pg/mL) to highly elevated levels (several tens of ng/mL). Nevertheless, integrating CL technology into a POCT platform poses significant challenges. Traditional CL assays rely on precise liquid handling, multiple washing steps, and a large/expensive imaging system, which complicate miniaturization efforts. As a result, most CL-based assay platforms have remained in centralized laboratories where the technical demands can be fully met. Efforts to incorporate CL into POCT systems have also faced limitations, including reliance on single-molecule enzyme conjugates with limited sensitivity,[30, 39, 41, 43] dependence on benchtop imaging stations that constrain portability and affordability,[30, 32, 36-39, 41, 42, 47] and inadequate performance of smartphone-based readers for CL imaging due to their optical/software limitations. To address these issues, advances in conjugate sensitivity, reader design, and assay platform miniaturization are required. This will ensure the implementation of a clinical laboratory-grade hs-cTnI assay in a fully portable and accessible format.

Here, we present a paper-based CL vertical flow assay (CL-VFA) for highly sensitive, precise, rapid, and affordable cTnI quantification, achieving clinically relevant sensitivity and precision comparable to laboratory-based standard hs-cTnI assays. The CL-VFA system integrates several innovations designed to address the critical limitations of existing CL-based POCT assays, (see **Figure 1b–f**): (i) a polymerized enzyme-based conjugate for enhanced CL signal intensity and sensitivity (see **Figure 2a**), (ii) a cost-effective, portable Raspberry Pi-based CL reader with robust imaging capabilities, eliminating the need for a complex benchtop readout system (see **Figure 1b** and **d**), (iii) a user-friendly, streamlined tray-based VFA cartridge that ensures stable and consistent CL imaging (see **Figure 1c** and **e**), and (iv) a neural network-driven computational pipeline for accurate classification and quantification of cTnI levels (see **Figure 1f**). This platform measures cTnI from a 50 µL serum sample within 25 min per test, combining the assay performance of a laboratory-based hs-cTnI analyzer with the simplicity of point-of-care diagnostics. For validation, we assessed the detection sensitivity and precision using both control serum samples spiked with cTnI and clinical patient serum samples. The initial validation using cTnI-spiked human serum resulted in a detection limit of 0.16 pg/mL and demonstrated high reproducibility, with an average coefficient of variation (CV) of 5.4%. In a blinded test involving 66 clinical samples from 34 patients, the predicted cTnI concentrations from the computational CL-VFA system strongly correlated (Pearson's correlation coefficient [$r$] of 0.979) with the ground truth values measured by an FDA-cleared laboratory instrument, with an average CV of 14.3%. These findings highlight the transformative potential of this deep learning-enhanced CL-VFA platform, offering a rapid, low-cost, and high-performance solution for accurate cTnI testing at the point of care. This approach represents a significant step forward in making high-sensitivity cardiac diagnostics more accessible across diverse healthcare settings.

## 2. Results and Discussion

## 2.1. Design and workflow of the CL-VFA

Our CL-VFA platform includes modular top and bottom cases that can be joined to provide custom formats for the assay and CL imaging phases. This preserves the key features of traditional VFA systems,[40, 52-58] such as simplicity, ease of use, and rapid assay times, while enabling a CL-based high-sensitivity assay in the POCT platform.

A paper-based sensing membrane (12 mm × 12 mm), attached to a tray, is designed to be easily moved to the imaging setup right after the assay (see **Figure 1c**). The sensing membrane consists of hydrophilic compartments on the paper surrounded by hydrophobic wax barriers. These barriers help direct and concentrate fluid flow into the reaction zones. There are a total of nine reaction spots on the membrane: 2 are used for cTnI detection (coated with anti-cTnI capture antibodies), 1 serves as a positive control (coated with secondary antibodies that bind anti-cTnI detection antibodies), and 5 are negative controls (treated with buffer). The remaining central spot acts as a blank fluid channel to reduce non-specific interactions. To prevent potential interference between the cTnI testing spots and positive control during the assay and imaging stages, these spots are positioned at the farthest distance from each other on the membrane.

In our VFA system, the wicking from the absorption pads, combined with the capillary action of engineered paper layers in the top case, ensures the uniform distribution of the sample and reagents as they pass through the stacked layers of paper through the sensing layer (see **Figure S1**). The full assay process is completed within 25 min, comprising 10 min for the immunoassay, 10 min for washing, and 5 min for incubating the CL reaction and capturing the signal (see **Figure 1e**). For the immunoassay phase, a combination of a 1st top and bottom cases using a twisting lock is employed. After wetting the stacked membrane layers and the capture antibodies on the sensing layer by adding 200 µL of running buffer, a mixture of 50 µL serum sample and 50 µL detection conjugate is added. This allows the cTnI antigen to bind to the capture antibodies in a sandwich assay format (capture antibody–cTnI–detection antibody), resulting in the concentration of the detection antibody bound to the sensing membrane being dependent on the cTnI levels in the sample. After that, a running buffer (350 µL) is added to help the sample-detection conjugate mixture move downwards and wash away non-specific detection antibodies. During the washing phase, the top case is switched to a new one, and then an additional 500 µL of running buffer is added to thoroughly wash away any unbound detection conjugates and serum components from the sensor membrane. It is essential to prevent non-specific signals and potential interference during the sensitive CL reaction.

For the CL reaction and imaging phases, the sensing membrane tray is transferred to the CL reaction/imaging setup, utilizing the 2nd top/bottom case assembly. The twisting case assembly mechanism remains the same as in the previous phase, ensuring user convenience for operation. Upon adding CL reagent solution (440 µL) into the 2nd top case, the solution is rapidly delivered (<1 s) to the sensing membrane, where it uniformly incubates within a chamber that is enclosed by a support stage, foam tape gasket, and top case. Subsequently, the VFA cartridge is inserted into the reader (see **Figure 1d**) and incubated for 4 min to allow the CL reaction to reach saturation. The CL reaction involves luminol, an enhancer, and hydrogen peroxide ($H_2O_2$), catalyzed by horseradish peroxidase (HRP) enzymes present in the detection antibody conjugate. The blue visible light ($\lambda_{max}$ = 425 nm) generated by this reaction is transmitted through the transparent acrylic window in the 2nd top case and captured by the portable reader. The CL signal is imaged over 30 s, and the images are computationally analyzed via neural network-based analysis algorithms to determine the cTnI concentration in the sample (see **Figure 1f**). This setup provides stable CL reaction and imaging conditions, thereby enabling precise and reproducible detection of cTnI at the pg/mL level.

## 2.2 PolyHRP-based high-sensitivity detection conjugates for CL-VFA

To achieve high-sensitivity detection of cTnI using the VFA platform, we employed a detection conjugate consisting of 15 nm AuNPs conjugated with PolyHRP-Streptavidin (PolyHRP) and biotinylated antibodies (**Figure 2a**).[32] PolyHRP, an engineered polymer containing multiple HRP and streptavidin (StA) molecules attached to a hydrophilic backbone, greatly enhances the CL signal generation compared to conventional single-molecule HRP-based conjugates. We hypothesized that PolyHRP-based conjugates, with their higher enzyme density per conjugate, would provide a significant sensitivity advantage over traditional HRP conjugates, where only a limited number of enzymes can be bound to a single nanoparticle or antibody. By increasing the number of active enzymes per conjugate, we achieved a stronger CL signal, resulting in improved detection limits for cTnI.

The synthesis of the conjugate follows a two-step process. First, PolyHRP is adsorbed onto the surface of 15 nm AuNPs. In the second step, the AuNP-PolyHRP complex is coupled with a biotinylated detection antibody through the strong affinity binding (with a dissociation constant, $K_d \approx 10^{-15}$ M) between StA and biotin.[59] The remaining biotin binding sites in StA are then blocked with biotin-BSA, which is smaller in size than the biotinylated antibody. While PolyHRP has traditionally been used as a sole assay label in standard laboratory techniques like enzyme-linked immunosorbent assay and western blot, our conjugation strategy uniquely incorporates 15 nm AuNPs as the core material. This approach offers several key advantages: (i) a compact conjugate size suitable for paper-based assays, allowing for easy transport through the porous paper matrix; (ii) a larger surface area that promotes efficient PolyHRP adsorption; (iii) a simplified conjugate synthesis process that relies on centrifugal washing rather than dialysis, reducing both time and the risk of protein loss, and (iv) AuNP-binding based analysis for more effective quality control, enabling simple spectral scanning and absorption peak measurements to assess the conjugation results and concentration titers.

As shown in **Figure 2b**, the CL signal of AuNP-PolyHRP-antibody conjugate (detection cut-off: $6.1 \times 10^{-5}$ OD) demonstrated significantly superior analytical performance compared to both its native colorimetric signal and the CL signal of the standard HRP conjugate (AuNP-standard HRP-antibody). Compared to the conventional colorimetric signal from AuNPs (detection cut-off: $2.1 \times 10^{-1}$ OD), the CL signal of AuNP-PolyHRP-antibody conjugate achieved a ~3400-fold increase in detection sensitivity. Similarly, it exhibited a **~210-fold** improvement over the CL signal of the standard HRP conjugate (detection cut-off: $1.3 \times 10^{-2}$ OD), supporting our hypothesis.

Although the AuNP conjugate has a relatively small size of approximately 60 nm,[32] optimizing its application in the VFA for cTnI detection required ensuring that it effectively passes through the top case's paper matrix and sensing membrane while minimizing non-specific signals. To achieve this, we systematically tested and optimized various assay conditions, including the assay protocol, running buffer composition, types of paper materials, and membrane-blocking reagents/concentrations. We identified three key factors that maximized the signal in positive samples (10 ng/mL of cTnI spiked in cTnI-free serum) while minimizing non-specific signals in negative samples (cTnI-free serum): (i) loading the serum–conjugate mixture in the 1st top case to increase immunoassay efficiency, and using a separate top case solely for the washing step (Test 1, **Figure S2**), which reduced non-specific binding and signal variations by effectively removing unbound conjugates, (ii) incorporating Triton X-100, a nonionic surfactant, into the running buffer (Test 2, **Figure S2**) to enhance flow rate and washing efficiency, and (iii) using a 0.45 µm pore size nitrocellulose (NC) membrane as the sensing membrane material (Test 3, **Figure S2**), which provided better permeability for the conjugate compared to the 0.22 µm membrane. These optimized conditions were established as the standard setting for the CL-VFA assay. The performance of the cTnI assay using the conjugate in combination with the portable reader will be discussed in the *High-sensitivity assay performance of the cTnI CL-VFA* section.

## 2.3. High-sensitivity CL imaging using a portable quantitative reader

The CL-VFA platform uses a custom-designed portable reader with a Raspberry Pi system (see **Figure 1d**). It incorporates cost-effective parts such as a CMOS camera module, lens, touchscreen display, 3D printed housing, and cartridge tray. Unlike previous colorimetric or fluorescent readers, which require reflectance imaging or fluorophore excitation through external light sources, the CL-VFA portable reader design benefits from the nature of CL, which generates light as a result of a chemical reaction with a high signal-to-noise (S/N) ratio. This eliminates the need for light-emitting diode-related components and simplifies the overall reader configuration, reducing complexity in the layout.

One key advantage of the CL modality in high-sensitivity biosensing is the ability to easily balance sensitivity and dynamic range by controlling exposure time and camera gain during the sensor readout. For example, longer exposure times can accumulate faint CL signals from low antigen concentrations, while shorter exposures help avoid signal saturation when measuring higher antigen concentrations, enabling accurate quantification over a larger dynamic range. Our graphical user interface (GUI) is designed to allow users to adjust the imaging exposure time (300 µs – 239 s) (see **Figure S3**), providing flexibility in optimizing sensitivity. Additionally, it supports time-lapse imaging by enabling the user to set both the time intervals between image capture and the total number of images, making it suitable for tracking CL signal dynamics over time. To minimize the impact of environmental light during these extended exposures, our reader uses an enclosed cartridge tray that offers dark conditions for imaging the VFA cartridge. Additionally, to mitigate heat generation from prolonged exposure, which could lead to interference or image distortion, a heatsink was integrated into the Raspberry Pi board and the CMOS image sensor, thus enabling passive but efficient thermal management throughout the readout process.

To assess the performance of our Raspberry Pi-based CL portable reader, we measured the S/N ratio of the CL signals obtained from the serially diluted/immobilized detection conjugate (AuNP-PolyHRP-antibody conjugate) on the sensing membrane. We compared the S/N ratio of the Raspberry Pi-based reader with a conventional benchtop system (cooled CCD-based) and a smartphone-based reader (CMOS-based, used for conventional colorimetric setting for VFA) for CL imaging (see **Figure 2c**). The Raspberry Pi-based CL portable reader demonstrated superior performance, achieving high sensitivity in detecting CL signals generated by the conjugate as low as $6.1\times10^{-5}$ OD. This sensitivity is 2.1-fold higher than the detection cut-off of the benchtop system ($1.3\times10^{-4}$ OD) and 26.2-fold higher than that of the smartphone-based reader ($1.6\times10^{-3}$ OD), highlighting its effectiveness in low-concentration analyte detection and suitability for high-sensitivity applications in point-of-care testing.

The Raspberry Pi-based CL portable reader has a compact design that places the sensing membrane at a distance of ~2 cm from the camera aperture, significantly closer than the tens of centimeters typical in benchtop systems (see **Figure 2d**). This proximity enhances our sensitivity by reducing light loss associated with greater distances, resulting in more efficient CL signal capture. As a result, minimized light loss and improved signal collection allowed the Raspberry Pi-based reader to achieve high sensitivity for detecting low concentrations of analytes, outperforming larger, more complex benchtop systems despite a compact, lightweight, and cost-effective design. In addition, the Raspberry Pi-based CL portable reader achieved a better S/N ratio compared to a smartphone-based reader with similar cost, size, and image sensor type. This improved performance can be attributed to several factors, such as enhanced sensor control options (i.e., optical and digital gain), larger pixel size (2.4 times larger), and better thermal management than the smartphone-based setup.

## 2.4. Tray-based assay cartridge for stable CL imaging

The tray-based assay cartridge was engineered to address signal stability issues observed when the 1st bottom case, initially used during the assay and washing steps, was reused for CL imaging. In the original setup, the sensing membrane was placed above stacked absorbent pads (see **Figure 2e**, left column), which worked well for the assay but led to unstable CL signals during imaging. Specifically, we found that the CL intensity failed to reach saturation and decreased over time with increased signal variation (average CV of 22.9%; see **Figure 2f**, w/ Absorption pad). We hypothesized that this issue arose because, despite the absorbent pads reaching their maximum capacity (~1 mL) during the assay and washing steps, evaporation through the ventilation hole continued, causing the CL reagent on the sensing membrane to continue to flow dynamically. As a result, the reagent was gradually reabsorbed into the pads, destabilizing the CL signal.

To validate this hypothesis, we performed an experiment where we carefully excised the sensing membrane from the absorbent-based bottom case following the assay/washing step and transferred it to a flat-bottom case without absorbent material (see **Figure S4a**). We then added the CL reagent, covered the membrane with a glass cover to prevent reagent flow or evaporation, and conducted CL imaging. Under these stabilized conditions, we observed that the CL signal reached saturation in 4 min with minimal signal deviation (average CV of 4.1%), confirming that stabilizing the reagent without flow on the sensing membrane is a critical parameter for consistent CL imaging (see **Figure S4b**).

To maintain the efficient assay/washing functionality of the original VFA cartridge while enabling stable CL imaging, we designed a tray-based interface that allows users to seamlessly transfer the sensing membrane for imaging. The new imaging cartridge incorporates plastic support instead of absorbent pads, preventing reagent movement and evaporation (see **Figure 2e**, right column). This setup preserves the original assay performance (see **Figure 2g**) while providing a stable, optimized environment for CL imaging (average CV of 4.6%; see **Figure 2f**, w/ Support stage). By combining the stability of the flat-bottom case conditions with enhanced usability, this tray-based cartridge enables hs-cTnI measurement in a user-friendly, practical format, as further described in the following section.

## 2.5. High-sensitivity assay performance of cTnI CL-VFA

The limit of detection (LoD) is a critical parameter for assessing the sensitivity of biomarker assays. To determine the LoD for our optimized CL-VFA platform, we conducted titration experiments using cTnI spiked in cTnI-free human serum. Serial dilutions were performed using the same serum medium, and the corresponding CL signals were captured using the portable reader. The signal intensity ($I_{Normalized}$) was calculated using the averaged test and negative control spot signals. The details are described in the *Computational analysis of CL-VFA signals* in the Methods section.

The detection performance of this system is illustrated in **Figure 3**. Using the Raspberry Pi-based portable reader, we acquired CL images at varying cTnI concentrations, resulting in apparent differences in concentration-dependent test spot signals (see **Figure 3a**). The calibration plot in **Figure 3b** demonstrates a strong correlation ($R^2 = 0.99$, $y=0.0147x^{0.4043}$) between the signal intensity and cTnI concentrations in the serum matrix, ranging from a few pg/mL to $10^5$ pg/mL (matching the clinically relevant range). The assay also exhibited excellent precision, showing an average CV of 5.4% between triplicate testing repeats of each sample. These results highlight both a broad sensing range of over 6 orders of magnitude and the reliability of this assay system.

To further assess the assay's sensitivity, we evaluated its performance in detecting lower cTnI concentrations (0–10 pg/mL) within the clinical cut-off level range (~40 pg/mL),[11] as shown in **Figure 3c**. The *P*-values derived from *t*-tests confirmed statistically significant differences ($P <0.05$) between the different cTnI levels under 10 pg/mL; for example, negative vs. 0.5 pg/mL ($P = 0.0019$) comparison underscores the platform's outstanding sensitivity and sensing resolution at or below the pg/mL level for high-sensitivity detection of cTnI.

Our analysis indicates that the limit of detection (LoD) for the cTnI CL-VFA is **0.16 pg/mL**, defined as LoD = Limit of Blank (LoB) + 1.645 × Standard Deviation (SD) of the lowest cTnI measurement.[48, 60] LoB is calculated as the Mean blank value + 1.645 × SD of the blank.[48, 60] The values measured were a mean blank of 0.0046, an SD of the blank equal to 0.00071, and an SD for the lowest cTnI concentration of 0.00073, with this lowest concentration being 0.5 pg/mL. The LoB was estimated to be 0.101 pg/mL by converting the CL intensity (y-value) into concentration using the optimal fitting curve shown in **Figure 3b** ($y=0.0147x^{0.4043}$). Subsequently, the LoD calculation resulted in 0.16 pg/mL through the same calibration curve.

These findings, along with the CL-VFA platform's performance, meet two critical criteria set by clinical guidelines for hs-cTnI testing: (i) a CV of ≤10% at the clinical cut-off, namely 99th percentile upper reference limit, and (ii) the capacity to identify cTnI levels at or over the assay's LoD in more than 50% of healthy patients (i.e., ~ 40 pg/mL).[11] This confirms that our methodology conforms to established clinical standards for hs-cTnI assays.

We measured 72 serum samples obtained from patients tested for cTnI levels at UCLA Health for practical validation using the CL-VFA platform (see **Figures 3d** and **3e**). The ground truth cTnI concentrations for these samples were measured and provided by UCLA Health using an FDA-cleared benchtop analyzer, enabling a hs-cTnI assay. This clinically used benchtop system has an analytical cut-off value of 4 pg/mL, where samples with cTnI levels above 4 pg/mL are quantified, and those below this level are reported as <4 pg/mL. In the following sub-section, these ground truth levels will be compared to the cTnI concentrations inferred by the neural network-based analysis approach of our system, allowing us to assess the accuracy and reliability of the CL-VFA platform in detecting clinically relevant cTnI levels.

The assay plots in **Figures 3d** and **3e** demonstrate a strong correlation ($R^2 = 0.99$, $y=0.0048x^{0.5}$) between the ground truth cTnI concentrations and CL intensity measured by the CL-VFA system across the entire quantifiable concentration range (≥4 pg/mL), with an average CV of 6.7%, demonstrating consistent inter-sensor measurement repeatability. Although the ground truth values of several samples were labeled as <4 pg/mL, we were able to detect minute differences in CL intensities between these samples using the CL-VFA platform, revealing subtle variations in cTnI levels that were otherwise indistinguishable by the benchtop analyzer. During the analysis, three samples within the $10^2$–$10^3$ ground truth level range were identified as outliers, as indicated in **Figure 3d**. Outliers were defined as samples displaying a signal intensity difference corresponding to a concentration deviation of at least one order of magnitude from the expected trendline. These outliers and three randomly selected normal samples that followed the expected trend line were reanalyzed using the same clinical benchtop analyzer to ensure accurate validation and inquire if samples degraded over time after their initial clinical measurements. Upon revalidation (see **Figure S5**), normal samples exhibited minimal changes in cTnI levels, ranging from 3.5% to 11%, consistent with their initial measurements. In contrast, outlier sample 1 showed an extreme degradation rate exceeding 99%, with its concentration decreasing to <2 pg/mL from an initial 512 pg/mL, which agrees with the <4 pg/mL intensity range detected by the CL-VFA— strongly indicating significant degradation of the sample. Outlier sample 2 exhibited a distinct discrepancy: despite showing a moderate degradation rate of 13.2%, its CL-VFA intensity fell within the <4 pg/mL range. Notably, the serum appeared darker red and more turbid compared to other samples in the same collection batch (typically yellowish), suggesting elevated levels of hemolysis and lipemia.[61, 62] These characteristics may have introduced assay interference, with a possible additional factor being the presence of autoantibodies against cTnI, which may mask target epitopes and interfere with the assay's ability to detect cTnI.[61, 62] Outlier sample 3 presented a more complex case, displaying characteristics seen in both samples 1 and 2. Outlier sample 3 showed a substantial degradation rate of 40%, indicating a notable decrease in troponin concentration. However, as with sample 2, additional assay interferents likely affected the VFA reading, categorizing it within the <4 pg/mL range.

These findings suggest that these 3 outlier samples, affected by either excessive degradation or potential assay interference, showed cTnI measurement inconsistencies compared to the stable readings in normal samples. Due to these discrepancies, the outliers were excluded from further neural network analysis to maintain the reliability of the model's predictions. We then utilized the trained neural network models on the validated CL-VFA signals to accurately quantify cTnI levels, as discussed in the next subsection.

## 2.6. Neural network-based cTnI quantification using CL-VFA data

Deep learning and neural networks enhance the performance of point-of-care sensors by effectively learning complex relationships between sensor output patterns and diagnostic outcomes, outperforming standard rule-based methods (e.g., linear regression, power fitting), which lack the complexity needed to approximate intricate functional relationships. We developed a neural network-based pipeline to accurately quantify cTnI concentrations in clinical serum samples using the signals generated by CL-VFA. This pipeline consists of four fully connected neural networks (see **Figure 4a**): one for classification ($DNN_{Classification}$) and three for biomarker quantification ($DNN_{Q<40}$, $DNN_{Q40-1000}$, and $DNN_{Q>1000}$). These networks work together to quantify cTnI over a large dynamic range of our clinical sample cohort, which spans from a few pg/mL to ~18,000 pg/mL, as shown in **Figure 3d**.

First, the $DNN_{Classification}$ network classifies the sample of interest into three cTnI concentration ranges: <40 pg/mL, 40–1000 pg/mL, and >1000 pg/mL. This classification is based on the averaged test, negative control, and positive control signals (inputs) generated by the CL-VFA. Following this initial classification, the sample of interest is processed by one of the three cTnI quantification networks—$DNN_{Q<40}$, $DNN_{Q40-1000}$, or $DNN_{Q>1000}$—each independently optimized and trained to accurately quantify cTnI concentrations within their respective concentration ranges. In the <40 pg/mL ($DNN_{Q<40}$) and 40–1000 pg/mL ($DNN_{Q40-1000}$) categories, these networks utilize the normalized CL signal ($I_{Normalized}$) as input, while the >1000 pg/mL network ($DNN_{Q>1000}$) takes the averaged test, negative control, and positive control signals for quantification (see *Computational analysis of CL-VFA signals* in Methods section for details). $I_{Normalized}$ incorporates test and negative control spots; however, for higher concentration ranges (i.e., >1000 pg/mL), the test spots exhibit more interference with negative controls due to the strong CL intensity, which causes the light to spread and reach adjacent areas. To account for this, $DNN_{Classification}$ and $DNN_{Q>1000}$ also take positive control spot signals for additional signal normalization.

In case of discrepancy between the concentration prediction from the quantification network and the output from $DNN_{Classification}$ (i.e., if the predicted concentration falls more than 10% outside the borders of the range predicted by $DNN_{Classification}$), the sample is marked as "*indeterminate*" due to the conflicting decisions of successive neural networks, and no cTnI concentration inference is performed. Quantification occurs only when predictions from the $DNN_{Classification}$ and the corresponding cTnI quantification network model align with each other. As a result, cTnI quantification measurement is done through this collaboration between the four neural network models (see *Neural network-based cTnI quantification pipeline* in Methods for details about these models).

Each of the four neural network models in the cTnI quantification pipeline was individually trained and optimized using distinct training and validation datasets tailored to their target cTnI concentration ranges. When tested on the validation set, $DNN_{Classification}$ achieved 93.5% accuracy, and the optimized quantification models resulted in a strong correlation (Pearson's *r* of 0.993) with the ground truth values from an FDA-approved clinical laboratory analyzer, with an average CV of <10%, demonstrating the reliable performance of all four models and good inter-sensor repeatability (see **Figure S6**). Additional details about the models' performance on the validation set are available in the *Neural network-based cTnI quantification pipeline* sub-section of the Methods section.

Next, this neural networks-based pipeline was blindly tested using 66 new serum samples from 34 patients, none of which were used during the training and optimization phases (**Table S2**). As shown in **Figure 4b**, DNN$_{Classification}$ achieved 95.5% accuracy on this blind testing set, correctly classifying all samples in the 40–1000 pg/mL and >1000 pg/mL ranges. However, three samples with ground truth cTnI concentrations of 32 pg/mL (2 samples) and 37 pg/mL (1 sample) were misclassified from the <40 pg/mL range into the 40–1000 pg/mL range.

Following the classification stage, samples from the <40 pg/mL, 40–1000 pg/mL, and >1000 pg/mL ranges were further quantified by the respective quantification models: DNN$_{Q<40}$, DNN$_{Q40-1000}$, and DNN$_{Q>1000}$. The combined cTnI concentration predictions from these models showed a good match with the ground truth cTnI levels from an FDA-approved analyzer, with a Pearson's $r$ of 0.984 (see **Figure 4c** and **Table S2**). In addition, cTnI concentration inferences showed good precision with an average CV of 14.3% between duplicate tests of patient samples. Some of the samples quantified by DNN$_{Q<40}$ had ground truth cTnI concentrations below 4 pg/mL and did not have quantitative concentration labels due to the cut-off level of the FDA-approved clinical analyzer. Quantitative predictions for such samples were plotted within the <4 pg/mL region of the ground truth axis (i.e., x-axis), as shown in **Figure 4c**. No samples in the blind testing set were labeled as "indeterminate" as all quantification predictions aligned with the classification network results. However, three samples with ground truth concentrations in the <40 pg/mL range were misclassified into the 40–1000 pg/mL range by DNN$_{Classification}$, which were successively quantified by DNN$_{Q40-1000}$, resulting in the predicted cTnI concentrations of 71 pg/mL, 105 pg/mL, and 72 pg/mL; for these samples, the ground truth concentrations were 32 pg/mL (2 samples) and 37 pg/mL (1 sample), respectively. Erroneous predictions by the classification and quantification models for these samples can be attributed to increased non-specific binding relative to the other clinical serum samples, which elevated the test spot signals, creating confusion with samples of higher concentrations.

We note that the decision to implement a multi-network architecture for the cTnI quantification pipeline, incorporating four neural networks, was based on evaluating several strategies with varying numbers of models (see **Table S3** for a summary of the comparison). For example, a single quantification model proved inadequate for reliably quantifying cTnI concentrations over the entire clinically relevant range due to the large dynamic range of cTnI levels observed in patient samples (from a few pg/mL to over $10^4$ pg/mL) and the limited number of training samples (see **Figure S7**). Therefore, to improve performance, we switched to a cascaded approach, where all samples are first classified into narrower concentration subranges, and each subrange is further quantified by a dedicated cTnI quantification model. We compared two configurations: one using three models (1 classification and 2 quantification) and another using four models (1 classification and 3 quantification). Ultimately, as shown in **Figure 4** and **Figure S8**, the four-model configuration was selected for its superior quantification accuracy (Pearson's $r$ of 0.984) compared to the three-model configuration (Pearson's $r$ of 0.846). Additionally, this four-model configuration can minimize the risk of overfitting, which can occur with limited data, while also increasing flexibility for incorporating new data in the future. The borders for the concentration subranges were selected to ensure a balanced representation of the samples within all ranges.

Importantly, our results highlight that systematic error mitigation steps, such as digital signal averaging and spot quality control, are critical for enhancing the analysis performance of CL-VFA. For example, averaging signals across similar spots within the test, negative control, and positive control conditions improved quantification precision, minimizing variability between individual spots caused by sensor fabrication errors, nonuniform flow, or CL signal leakage between adjacent spots. Using single-spot inputs instead of averaged signals increased the CV from 14.3% to 16.7% across the quantification range, with a notable rise in the lower concentration range (<1000 pg/mL), where the CV increased from 15.3% to ~20%, indicating reduced precision (see **Figure S9** and **Table S3**). In addition, negative control spots underwent a digital quality control step, which excluded individual spots (among the five original spots) from averaging if they fell outside the 95% confidence interval (see *Computational analysis of CL-VFA signals* in the Methods section for more details). Omitting this quality control step lowered the classification accuracy of DNN$_{Classification}$ to 94%, primarily due to the additional misclassification of one borderline sample from the 40–1000 pg/mL range (with a ground truth cTnI concentration of 51 pg/mL) into the <40 pg/mL range (see **Figure S10**). Therefore, incorporating signal averaging and digital quality control enhanced the robustness of CL-VFA, ensuring accurate diagnostics outcomes despite the low-cost nature of the VFA platform.

We also note that neural network-based cTnI quantification outperforms conventional deterministic methods in accuracy and precision. In our case, the optimal rule-based method consists of three independently optimized power-fitting curves, each tailored to a specific cTnI concentration range (<40 pg/mL, 40–1000 pg/mL, or >1000 pg/mL). For a fair comparison, these power-fitting models were trained using the same dataset as the quantification neural networks and were blindly tested on identical blind testing sets. As shown in **Figure S11**, cTnI concentration predictions by the rule-based models revealed a weaker correlation with the ground truth concentrations (Pearson's *r* of 0.912) compared to the optimal neural network models (Pearson's *r* of 0.984). Moreover, the CV between the duplicate tests was >21% for the power-fitting models, substantially higher than that observed with the neural networks. This inferior performance likely stems from the simplicity of the power-fitting models, which lack the large number of trainable parameters available in shallow neural networks, limiting their ability to capture complex functional relationships between input signals and target cTnI concentrations.

Similarly, DNN$_{Classification}$ outperforms other machine learning techniques, such as random forests and logistic regression. For instance, on the same blind testing set, the random forest model achieved only 91.0% accuracy, showing inferior performance (see **Figure S12a**). Furthermore, while the logistic regression model matched the overall accuracy of DNN$_{Classification}$, it misclassified some samples from the 40–1000 pg/mL range into the <40 pg/mL range, generating false negative outcomes that could overlook an ongoing MI (see **Figure S12b**). Such errors pose a greater health risk to patients than false positives when samples with <40 pg/mL ground truth are classified into higher ranges. DNN$_{Classification}$ demonstrated superior accuracy for higher concentration samples, correctly classifying all samples in the 40–1000 pg/mL and >1000 pg/mL ranges and mitigating false negative predictions. Therefore, we selected the DNN$_{Classification}$ model for the classification step of our neural network-based cTnI quantification pipeline.

In this study, using a hs-cTnI clinical assay for the ground truth measurements allowed us to obtain accurate cTnI levels even below 40 pg/mL, enabling neural network-based quantification within the 4–40 pg/mL range. This enabled our POC platform to accurately quantify cTnI concentrations below the 40 pg/mL threshold, which is a commonly recognized cut-off for myocardial infarction risk assessment.[10, 11] By facilitating reliable detection at these lower levels, the CL-VFA platform has the potential to improve early cardiac event detection and monitoring. For example, it could shorten the interval for confirmatory testing in suspected AMI patients, reducing the wait from 2–3 hours after initial measurement to potentially under 1 hour,[63] for quicker observation of troponin level increases. Additionally, for patients without initial signs of elevated risk, it could support rapid discharge under the 0-hour rule-out strategy,[64] enhancing clinical efficiency and patient throughput.

## 2.7. Cost analysis

At the current laboratory scale, the cost per test for the CL-VFA assay, including reagents, paper materials, and the plastic cartridge, is estimated to be $4.25 (see **Table S4**). However, substantial cost reductions are expected with large-scale production practices. For example, cartridge housing, which currently represents about 32% of the total test cost, could transition from 3D printing to injection molding, bringing housing costs close to a few cents. Similarly, antibody, chemical reagent, and raw

material costs could be significantly reduced through economies of scale, potentially enabling a per-test cost of under $1–2. This scalability could provide a feasible pathway to making hs-cTnI testing more affordable, allowing for widespread clinical and point-of-care applications, especially in resource-limited settings.

Additionally, the cost-effectiveness of the Raspberry Pi-based portable reader further enhances the practical appeal of this system for point-of-care settings. Unlike traditional benchtop imaging systems that cost upwards of $30,000, the Raspberry Pi-based reader, combined with a streamlined CL-VFA cartridge, offers a robust yet affordable alternative (see **Figure 2d**). The reader uses cost-effective components, resulting in a total system cost of ~$222 (see **Table S4**). This affordability does not compromise performance, making it suitable for high-sensitivity diagnostic applications, even in resource-limited settings. The low production costs of the sensor and the reader suggest that this platform could be widely deployed for routine testing of cTnI, particularly in decentralized healthcare settings.

## 3. Conclusion

In this study, we developed a neural network-enhanced paper-based high-sensitivity CL-VFA for quantitative cTnI detection. The CL-VFA enabled high sensitivity detection of cTnI (LoD of 0.16 pg/mL) with good precision (average CV of 14.3%, over the entire quantification range) within 25 min per test using human serum samples. Our design integrates multiple innovative components to achieve hs-cTnI detection, including (i) a PolyHRP-based AuNP detection conjugates that improve detection sensitivity by more than 2 orders of magnitude over conventional HRP-based CL methods; (ii) a cost-effective, handheld Raspberry Pi-based portable CL reader that outperforms a benchtop CL imaging station despite having a significant cost reduction; (iii) a streamlined tray-based VFA assay cartridge that ensures user-friendly assay operation and stable CL imaging; and (iv) a robust neural network-based concentration inference algorithm that leverages deep learning for accurate quantification across a wide dynamic range (over 6 orders of magnitude), enhancing both the precision and reliability of cTnI measurements.

Compared to both commercial benchtop and point-of-care cTnI assays, as well as recent advancements reported in the literature using CL-related sensing modalities, our CL-VFA platform demonstrates competitive performance in sensitivity, dynamic range, precision, and portability (see **Table 1** and **Table S1**) surpassing traditional benchtop analyzers by an order of magnitude in sensitivity. This work paves the way for low-cost, high-performance point-of-care diagnostics, addressing the critical need for highly sensitive cardiac biomarker testing that has traditionally been limited to central laboratory-based settings. Furthermore, the CL-VFA platform holds promise for broad applications across various biomarkers beyond cTnI, underscoring its adaptability and impact. Looking forward, leveraging the inherent multiplexing capabilities of the VFA, multiplexed CL-VFAs capable of simultaneously detecting cTnI and other key cardiovascular disease biomarkers can be developed. Such an enhancement would enable comprehensive cardiovascular risk assessment in a single test, improving diagnostic precision and patient outcomes. Additionally, integrating rapid plasma or serum extraction technology for whole-blood testing will support the streamlined implementation of CL-VFA in distributed clinics or other point-of-care setups. This advancement would further extend the CL-VFA's practical applications, facilitating rapid and reliable POCT in various healthcare environments. We envision that the versatile potential of the CL-VFA would make a significant step toward more accessible, efficient, and comprehensive diagnostics across various healthcare settings, driving forward the democratization of advanced healthcare.

## 4. Methods

***Conjugate preparation***: The conjugation of PolyHRP-Streptavidin (PolyHRP) and biotinylated antibody to 15 nm AuNP was performed through adsorption and affinity binding, respectively. In the first step, for the AuNP conjugation with PolyHRP, 15 nm AuNP (1 mL; BBI Solutions) was mixed with 100 mM borate buffer (100 µL, pH 8.5; Thermo Scientific) and stabilized for 5 min at room temperature (RT). Subsequently, PolyHRP (11 µL, 1 mg/mL; streptavidin-peroxidase polymer, Sigma) was added to the AuNP mixture, followed by incubation for 1 h at RT using a rotary mixer (20 rpm). After incubation, bovine serum albumin (BSA; 10 µL, 10% w/w; Thermo Scientific) was added to block the AuNP surface, and the mixture was incubated for 1 h at RT. The conjugate was then washed three times via centrifugation (25,000g, 30 min, 4 °C), using 10 mM borate buffer (pH 8.5) as the washing medium. After the washing steps, the conjugate pellet was resuspended in 10 mM PBS (100 µL, pH 7.2; Thermo Scientific) for storage as a 10× concentrated conjugate solution, which was subsequently transferred to a new reaction tube.

In the second step, the conjugation of AuNP-PolyHRP with a biotinylated antibody was performed. Biotin was labeled to the anti-cTnI detection antibody (19C7, Hytest) following the manufacturer's protocol (EZ-Link™ Sulfo NHS-LC-LC-Biotin, Thermo Scientific), and the concentration of the purified biotinylated antibody was quantified using a Nanodrop spectrophotometer (Thermo Scientific). Briefly, 12 µg of biotinylated antibody was added to the AuNP-PolyHRP solution (e.g., the addition of 4.8 µL of 2.5 mg/mL biotinylated antibody), followed by gentle mixing and incubation for 2 h at RT using an orbital shaker (55 rpm). Next, 5 µL of biotin-BSA (5 mg/mL; Thermo Scientific) was added and incubated for 1 h at RT (at 55 rpm on an orbital shaker) to block any remaining biotin binding sites on the streptavidin. Finally, the conjugate was washed three times with 10 mM borate buffer (pH 8.5) by centrifugation (25,000g, 30 min, 4 °C) to remove unbound components. The final pellet was resuspended in 100 µL of storage buffer containing 10 mM Fe(II)-EDTA, 4% (w/w) trehalose, 0.1% (w/w) BSA, and 1% (v/v) Triton X-100 in PBS (10 mM, pH 7.2).[32] Absorption spectra and conjugate concentration were analyzed using a microplate reader (Synergy H1; BioTek), resulting in a redshift of the peak wavelength after conjugation (from 520 nm to 526 nm, see **Figure S13**). Conjugates were stored at 4 °C at 8 OD concentrations until use. For preparing the 15 nm AuNP-standard HRP conjugate, we used StA-peroxidase conjugate (Sigma) instead of PolyHRP, but the conjugation process remains the same.

***Preparation of sensing membrane and assembly of assay cartridges***: The sensing membrane was fabricated following a sequential process that included wax printing, heat treatment, antibody deposition, and blocking. Initially, a wax printer (Xerox) was used to print nine active zones (eight reaction areas plus a central flow zone) on an NC membrane (0.45 µm, Sartorius), surrounded by a black background. These wax-printed NC membranes were then baked at 120 °C for 55 s in a forced-air convection oven (Across International), creating defined compartments for reaction sites and fluid flow paths. Post-heat treatment, the waxed areas turned hydrophobic, blocking the NC membrane pores, while the unwaxed regions remained hydrophilic to allow samples and aqueous solutions to flow. Up to 30 sensing membranes were arranged in a 6 × 5 grid with 1 mm spacing and processed simultaneously within a single batch.

To apply antibodies, anti-cTnI capture antibody (0.8 µL, 1 mg/mL; 560, Hytest), goat anti-mouse IgG (0.8 µL, 20 µg/mL; Southern Biotech), and 10 mM PBS (0.8 µL, pH 7.2) were spotted onto the designated test, positive control, and negative control regions, respectively. Membranes were air-dried at 37 °C for 15 min before immersion in a 1% (w/v) BSA blocking solution for 30 min at room temperature (RT). Following blocking, the membranes were dried again at 37 °C for 15 min. Each large membrane sheet was then carefully divided into individual sensing membranes for the assay using a razor and attached to the sensing membrane tray with double-sided adhesive foam tape.

To assemble the CL-VFA, the paper materials were prepared according to established methods.[40, 52-56] Raw materials were precisely cut to the required dimensions using a $CO_2$ laser cutter (Trotec). For the 1st top (used for immunoassay and washing steps) assembly, the cartridge was constructed by layering several paper components, including an absorption layer, flow diffuser, primary spreading layer, interpad, secondary spreading layer, and supporting layer, which were bonded with adhesive foam tape for stability. Wax printing and baking processes similar to those used for the sensing membrane were applied to create concentric circular patterns on both the flow diffuser and the outer frame of the supporting layer. To minimize non-specific binding, the flow diffuser, interpad, and supporting layers were treated with a 1% (w/v) BSA solution. For the bottom case, five absorption pads were positioned at the cartridge's center, and the assembled trays were stacked, as shown in **Figure S1**.

*CL-VFA cartridges*: The sensing membrane tray was 3D-printed using an Ultimaker S3 (PETG filament, Ultimaker) with a layer thickness of 150 µm. The top and bottom cases for the CL-VFA were fabricated using a Form 3 printer (Formlabs) with gray resin at a resolution of 100 µm. When the bottom case, sensing membrane tray, and 1st top case are assembled, the design compresses the paper layers to 25% of their original thickness, optimizing flow uniformity and enhancing both assay and washing efficiency.[40] For the CL reaction and imaging, the 2nd top case was created from a transparent acrylic sheet, laser-cut to precise dimensions (16.3 mm × 16.3 mm × 1 mm), and attached to the 3D-printed body using clear acrylic glue. The acrylic window includes a ventilation port (0.7 mm in diameter) positioned at the lower left corner, facilitating air displacement from the reaction chamber as CL reagent solutions are introduced. Upon assembly, the foam tape attaching the sensing membrane to the tray aligns with the extruded edges of the top case, creating a secure seal. This configuration effectively prevents leakage, evaporation, and displacement of the injected CL reagent solution, eliminating the need for an additional gasket. The reagent inlet of the 2nd top case is designed to hold up to 520 µL of solution.

*Raspberry Pi-based portable CL reader*: Our custom-designed CL reader used a Raspberry Pi (model 3b) board to control the camera module through a user-friendly GUI. The camera module consisted of a Raspberry Pi HQ camera (Adafruit) coupled to a C-mount lens (Adafruit), enabling the capture of CL reactions over long exposure times of up to 239 s, achieving high sensitivity. The camera was connected to Raspberry Pi via a dedicated camera serial interface inlet on the board, eliminating the need for additional software drivers. The GUI displayed a real-time camera feed and allowed users to set the exposure time, number of time-lapse images, and time intervals between subsequent images (see **Figure S3**). The reader was assembled with a Raspberry Pi board, camera module, heatsink, and touchscreen display (Elecrow), enclosed in a 3D-printed case, consisting of the housing, camera holder, and cassette tray. The 3D-printed parts were produced by Object 30 (Stratasys) 3D printer. The reader-supported pedestal installation used optical posts, allowing both handheld and benchtop use (see **Figure 1d**).

*Assay operation*: The assay operation involves two primary stages: the immunoassay and the CL reaction/imaging. For the immunoassay stage, the device is assembled by combining the 1st bottom case, the sensing membrane tray, and the 1st top case. Initially, the device is activated by introducing 200 µL of running buffer containing 1% (v/v) Tween-20, 0.5% (v/v) Triton X-100, and 1% (v/v) BSA in 10 mм PBS (pH 7.2). After 30 s, when the buffer is fully absorbed, a mixture of 50 µL of serum sample and 50 µL of conjugate (1 OD) is applied and allowed to incubate for 1 minute. Following this, 350 µL of running buffer is added to facilitate the flow of solutions through the CL-VFA device, removing any unbound conjugates and target molecules from the sensing membrane to minimize non-specific binding. After 8.5 min, the 1st top case is removed and replaced with a new one. An additional 550 µL of running buffer is added for a secondary washing step, which takes 10 min to complete. After the washing step is complete, the sensing membrane tray is transferred to the 2nd bottom case, which includes a support stage, and then assembled with the 2nd top case. Next, 440 µL of CL reagent solution (Thermo Scientific) is added to the cartridge. The cartridge is then placed into a portable reader for 4 min to allow the CL signal to reach saturation. Imaging is then performed for 30 seconds to capture the CL signal for analysis.

Through extensive timed experiments with over 300 activated cartridges, the total assay time was established at 25 min. Each step was optimized, with the immunoassay incubation period tailored for complete binding and efficient washing within 20 min, followed by a 5-min phase dedicated to the CL reaction and signal acquisition. Computational analysis was completed in less than 1 s per sample, with trained neural network models inferring cTnI concentrations in under 0.5 s, a negligible duration relative to the overall assay time.

*cTnI spiked and clinical serum samples*: To optimize and validate the assay, cTnI-spiked serum samples were prepared by adding human heart-derived cTnI standard antigen (I-T-C complex; Lee Biosolutions) into cTnI-free serum (Hytest). Serial dilutions were performed with the same serum to achieve a range of cTnI concentrations. The cTnI antigen was dispensed in 1 µL aliquots and stored at -80°C for stability. Fresh aliquots were thawed and used immediately before each experiment to avoid degradation. Clinical serum samples, obtained from UCLA Health under IRB protocol # 20-002084, involved remnant/existing specimens collected separately from this study. The ground truth cTnI levels in clinical samples were measured shortly after sample collection using a hs-cTnI assay (Access 2; Beckman Coulter) at UCLA Health. This clinical analyzer has a detection threshold of 4 pg/mL and reports values below this threshold as <4 pg/mL while accurately quantifying levels ≥4 pg/mL. In total, 72 clinical serum samples were analyzed with the CL-VFA, encompassing 58 samples with cTnI concentrations of 4 pg/mL or above, and 14 samples with levels below 4 pg/mL (see **Table S2**). The clinical samples were stored at -80°C and thawed at 4°C prior to testing.

*Computational analysis of CL-VFA signals*: Immunoreaction spots were segmented from the captured sensing membrane images using ImageJ software, and the same type of spots within a test, negative control and positive control conditions were averaged, generating 3 output signals per CL-VFA:

$$I_{Raw} = [\overline{X}_{Test}, \overline{X}_{(-)}, X_{(+)}], \tag{1}$$

where $\overline{X}_{Test}$ and $\overline{X}_{(-)}$ refer to the averaged test and negative control signals, respectively. $X_{(+)}$ refers to the positive control signal. Normalized signals were calculated as:

$$I_{Normalized} = 1 - \frac{2^{16}-1-\overline{X}_{Test}}{2^{16}-1-\overline{X}_{(-)}}. \tag{2}$$

Before any averaging, negative control spots underwent digital quality control, eliminating any spots outside the 95% confidence interval. Spots were excluded from $\overline{X}_{(-)}$ based on the 95% acceptance range from the statistical distribution of all negative control spots from the CL-VFAs activated during the clinical study, excluding the outliers reported in the Results section, i.e.:

$$X_{(-)}^{Low} < X_{(-),i} < X_{(-)}^{High}, i = 1 \ldots N_{Neg}, \tag{3}$$

where $X_{(-)}^{Low}$ and $X_{(-)}^{High}$ are calculated as ±1.96 SD from the mean of the distribution of all negative control spots from the clinical study. $X_{(-),i}$ denotes the signal of the negative control spot with index i and $N_{Neg} = 5$ is the total number of negative control spots per CL-VFA. $\overline{X}_{(-)}$ is calculated over the negative control spots that remained valid after this digital quality control step. If all negative control spots in a CL-VFA sample were excluded, the entire sample was removed from the dataset. As a result, 4 CL-VFAs were excluded due to negative control failure, leaving 134 CL-VFAs, which were split between training, validation, and testing datasets for neural network-based cTnI quantification. Digital quality control was not applied to the test and positive control spots since each CL-VFA contained only 2 test spots and 1 positive control spot.

*Neural network-based cTnI quantification pipeline*: The neural network-based pipeline for quantification of cTnI in clinical samples consisted of four models, including one classification ($DNN_{Classification}$) and three quantification ($DNN_{Q<40}$, $DNN_{Q40-1000}$, $DNN_{Q>1000}$) models (**Figure 4a**). All models represented shallow fully connected neural networks, which were trained and optimized separately on different portions of the clinical dataset. Inputs to $DNN_{Classification}$ and $DNN_{Q>1000}$ represented averaged raw signals from the CL-VFA, i.e., $X_{IN} = I_{raw}$. For $DNN_{Q<40}$ and $DNN_{Q40-1000}$, input signals represented normalized intensities, i.e., $X_{IN} = I_{Normalized}$.

Input signals to all networks were standardized according to the equation below:

$$X_{IN}^{st} = \frac{X_{IN} - <X_{IN}>}{\sigma(X_{IN})}, \qquad (4)$$

where $<*>$ and $\sigma(*)$ denote elementwise mean and SD of the input signals, respectively, calculated over the training data.

$DNN_{Classification}$ contained 3 hidden layers (128, 64, 32 units), each with "ReLU" activations and L2 regularization ($\alpha$=1e-3). All hidden layers were complemented by a batch standardization layer and a 0.5 dropout. $DNN_{Classification}$ was compiled using a categorical cross-entropy loss function ($L_{CCE}$), a learning rate of 1e-3, and a batch size ($N_b$) of 5. $L_{CCE}$ is expressed by:

$$L_{CCE}(y, y') = -\frac{1}{N_b} \sum_{n=1}^{N_b} \sum_{x}^{C} y_{n,x} \log y'_{n,x} \qquad (5)$$

where $y_{n,x}$ represent the gold standard binary labels of the $C = 3$ concentration range classes per sample, i.e., $x \in \left\{ cTnI < 40 \frac{pg}{mL}; \; cTnI = 40 - 1000 \frac{pg}{mL}; \; cTnI > 1000 \frac{pg}{mL} \right\}$, and $y'_{n,x}$ are the model's predicted probabilities for these classes, calculated using the sigmoid activation function:

$$y'_{n,x} = \frac{1}{1 + e^{-\hat{y}_{n,x}}}, \qquad (6)$$

where $\hat{y}_{n,x}$ refer to the model's inference before the sigmoid function for the $C = 3$ cTnI concentration range classes. The final predicted cTnI concentration range class per sample was determined as an output node with the highest probability from the sigmoid activation.

$DNN_{Q<40}$ and $DNN_{Q40-1000}$ consisted of 3 hidden layers with 256, 128, and 64 units, while $DNN_{Q>1000}$ contained 3 hidden layers with 128, 64, and 32 units. All hidden layers in all three quantification networks had 'ReLU' activations and L2 regularization ($\alpha$=1e-3). Furthermore, all hidden layers in all three models were supplemented by batch standardization and a 0.25 dropout. All quantification models were trained with a mean squared logarithmic error (MSLE) loss function, a learning rate of 1e-3 and $N_b$ = 5. MSLE loss is expressed by:

$$MSLE = \frac{1}{N_b} \sum_{i=1}^{N_b} (log(y_i + 1) - log(y'_i + 1))^2, \qquad (7)$$

where $y_i$ are the gold standard cTnI concentrations and $y'_i$ are the model-predicted concentrations. The unit of ground truth concentrations for $DNN_{Q<40}$ and $DNN_{Q40-1000}$ was in pg/mL, while that of the ground truth concentrations for $DNN_{Q>1000}$ was in ng/mL. Ground truth labels for $DNN_{Q40-1000}$ were normalized by 40 pg/mL. Configurations of the four models were optimized using a 4-fold cross-validation procedure on the validation sets.

$DNN_{Classification}$ was trained on 68 CL-VFA samples activated with serum samples from 35 different patients and validated on 62 CL-VFA samples from 32 patients. cTnI concentrations in the training and validation sets for $DNN_{Classification}$ varied between <4 pg/mL and ~10,000 pg/mL, covering typical clinical ranges of hs-cTnI testing. Optimized $DNN_{Classification}$ demonstrated 93.5% accuracy on the validation samples, correctly classifying all samples in <40 pg/mL and >1000 pg/mL ranges (see **Figure S6a**). Four out of 15 samples from 40–1000 pg/mL range were misclassified into the <40 pg/mL range. Ground truth cTnI concentrations in these samples were 46 pg/mL (2 samples), 56 pg/mL (1 sample), and 139 pg/mL (1 sample). Blind testing of $DNN_{Classification}$ on 66 additional serum samples from 34 patients showed 95.5% accuracy, as reported in the Results section (**Figure 4b**).

$DNN_{Q<40}$ was trained on 47 CL-VFA samples activated with serum samples from 24 different patients and validated on 34 CL-VFA samples activated with serum samples from 17 patients. The training dataset for $DNN_{Q<40}$ had cTnI concentrations between <4 pg/mL and ~100 pg/mL. Ground truth concentrations in samples from the <4 pg/mL range were assigned to 4 pg/mL, and MSLE for these samples only increased if the model prediction exceeded 4 pg/mL. $DNN_{Q40-1000}$ was trained on 18 samples from 10 patients and further validated on 15 samples from 8 patients. Training samples for $DNN_{Q40-1000}$ had cTnI concentrations from ~20 pg/mL to ~1,500 pg/mL. Finally, $DNN_{Q>1000}$ was trained on 19 samples from 10 patients and then validated on 13 samples from 7 patients. cTnI concentrations for $DNN_{Q>1000}$ varied between ~500 pg/mL and ~10,000 pg/mL. Validation samples for the three quantification models were selected from 62 validation samples used for $DNN_{Classification}$ according to the target quantification ranges (i.e., <40 pg/mL, 40–1000 pg/mL, or >1000 pg/mL, see **Table S2** for more details).

cTnI quantification was reported if and only if the concentration prediction by the quantification model aligned with the concentration range predicted by $DNN_{Classification}$, i.e., the predicted concentration value stayed within the range predicted by $DNN_{Classification}$ or fell within 10% from the range borders. Otherwise, the sample was labeled as "indeterminate" and eliminated from quantification. Based on this rule, two validation samples were labeled as "indeterminate". One sample from the validation set, with a ground truth concentration measurement of 139 pg/mL, was classified into the <40 pg/mL range by $DNN_{Classification}$; however, $DNN_{Q<40}$ predicted a concentration value of ~60 pg/mL, exceeding the 40 pg/mL border by 50%. Similarly, another serum sample with 1,613 pg/mL ground truth was correctly classified as >1000 pg/mL by $DNN_{Classification}$; however, prediction from $DNN_{Q>1000}$ was ~590 pg/mL, ~40% lower than the 1000 pg/mL threshold. These 2 samples were excluded from quantification and labeled as indeterminate due to the inconsistency between the quantification and classification neural networks; the quantification models were validated on 60 samples from 31 patients (see **Figure S6b**). None of the samples from the blind testing dataset had contradicting predictions between the classification and quantification models. Therefore, none of the blind testing samples were labeled as "indeterminate."

Predicted cTnI concentrations for the 60 validation samples demonstrated a good correlation with the gold standard values from an FDA-approved analyzer, with a Pearson's *r* of 0.993 and a CV of 8.7% between duplicate testing repeats (see **Figure S6b**). Blind testing of $DNN_{Q<40}$, $DNN_{Q40-1000}$, and $DNN_{Q>1000}$ on 66 new samples, which were not used in training and optimization, also demonstrated reliable quantitative performance, achieving a Pearson's *r* of 0.984 (**Figure 4c**). Clinical samples from the <4 pg/mL concentration range were excluded from the calculation of Pearson's *r* and CV for both validation and blind testing sets since they did not have quantitative ground truth concentration measurements. Further details about the quantification models' performance on the blind testing dataset are reported in the Results and Discussion section, under the "*Neural network-based cTnI quantification using CL-VFA data*" sub-section.

Training times for $DNN_{Classification}$, $DNN_{Q<40}$, $DNN_{Q40-1000}$, and $DNN_{Q>1000}$ were 6.4, 15.8, 14.6, and 17.0 min respectively. Blind testing times were considerably lower, not exceeding 0.5 s for any of the 4 models. The models were trained and tested on a desktop computer with a GeForce GTX 1080 Ti (NVIDIA). The neural network-based cTnI quantification pipeline was developed in Python using NumPy, TensorFlow, and Keras libraries.

*Statistical analysis*: For assay validation and tests involving cTnI-spiked serum samples, data were presented as the mean ± SD, derived from a minimum of three independent measurements. Results for clinical sample tests were reported as the mean of two replicate measurements ± SD. Detailed information on the number of experimental replicates is available in each figure caption. The CV was computed as the ratio of the SD to the mean, expressed as a percentage. Group differences were evaluated using an unpaired two-sample t-test, with statistical significance set at $P < 0.05$.

**Supplementary Information** includes **Table 1**.

# Figures

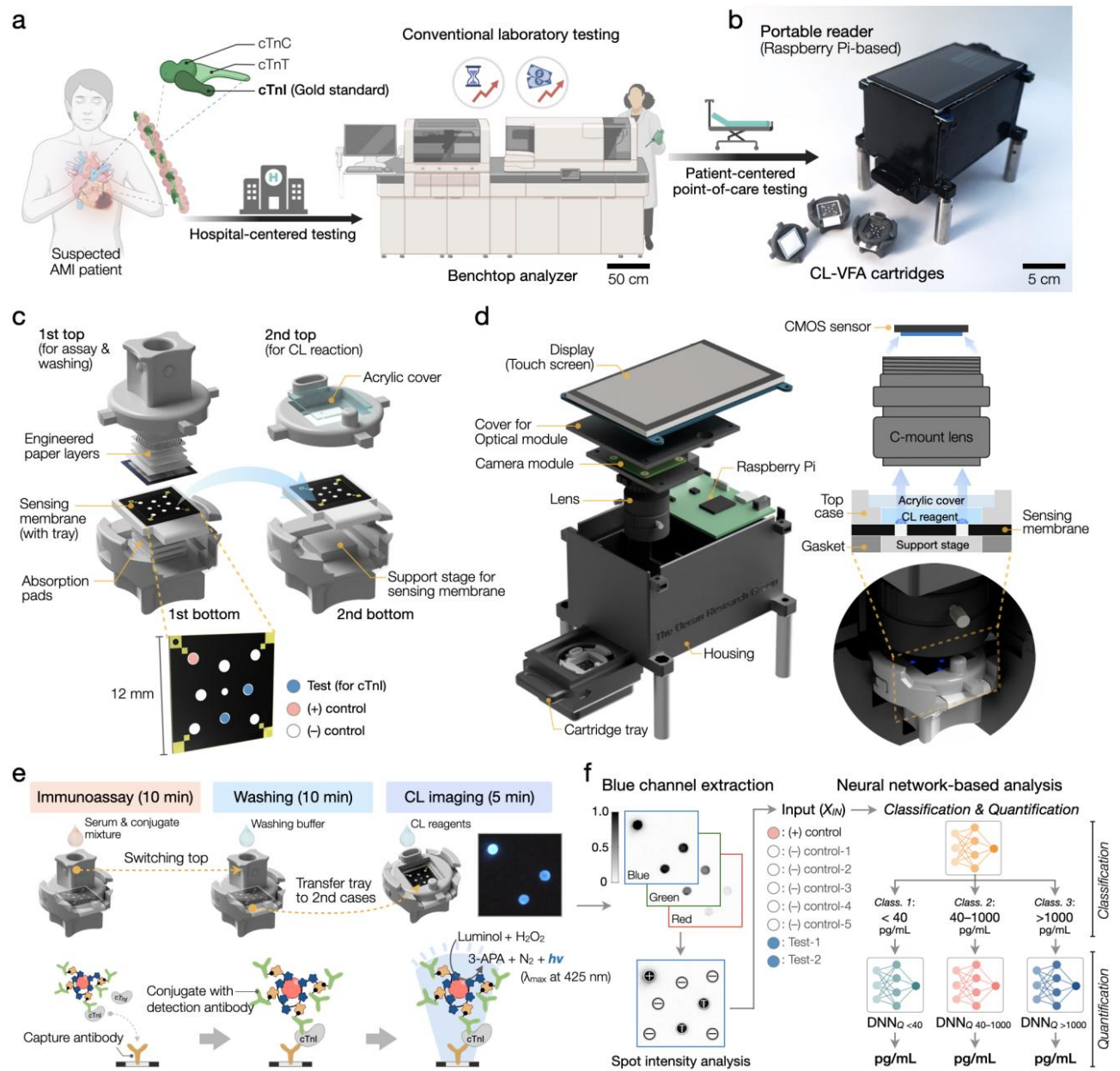

**Figure 1.** Overview of deep learning-enhanced paper-based Chemiluminescence (CL)-VFA for high-sensitivity cTnI testing in point-of-care settings. (a) The conventional hospital-centered approach for diagnosing AMI patients and its transition to a patient-centered POCT model. (b) Components of the CL-VFA system, including assay cartridges and a portable reader for rapid and accessible POCT. (c) Structural details of the CL-VFA; top and bottom cases and the transferable sensing membrane. (d) Exploded view of the portable reader assembly and internal configuration of the cartridge tray during the CL imaging process. (e) Assay workflow of CL-VFA with schemes of immunoassay and CL reaction steps. (f) Computational analysis workflow powered by deep learning, detailing the processing and analysis of assay data.

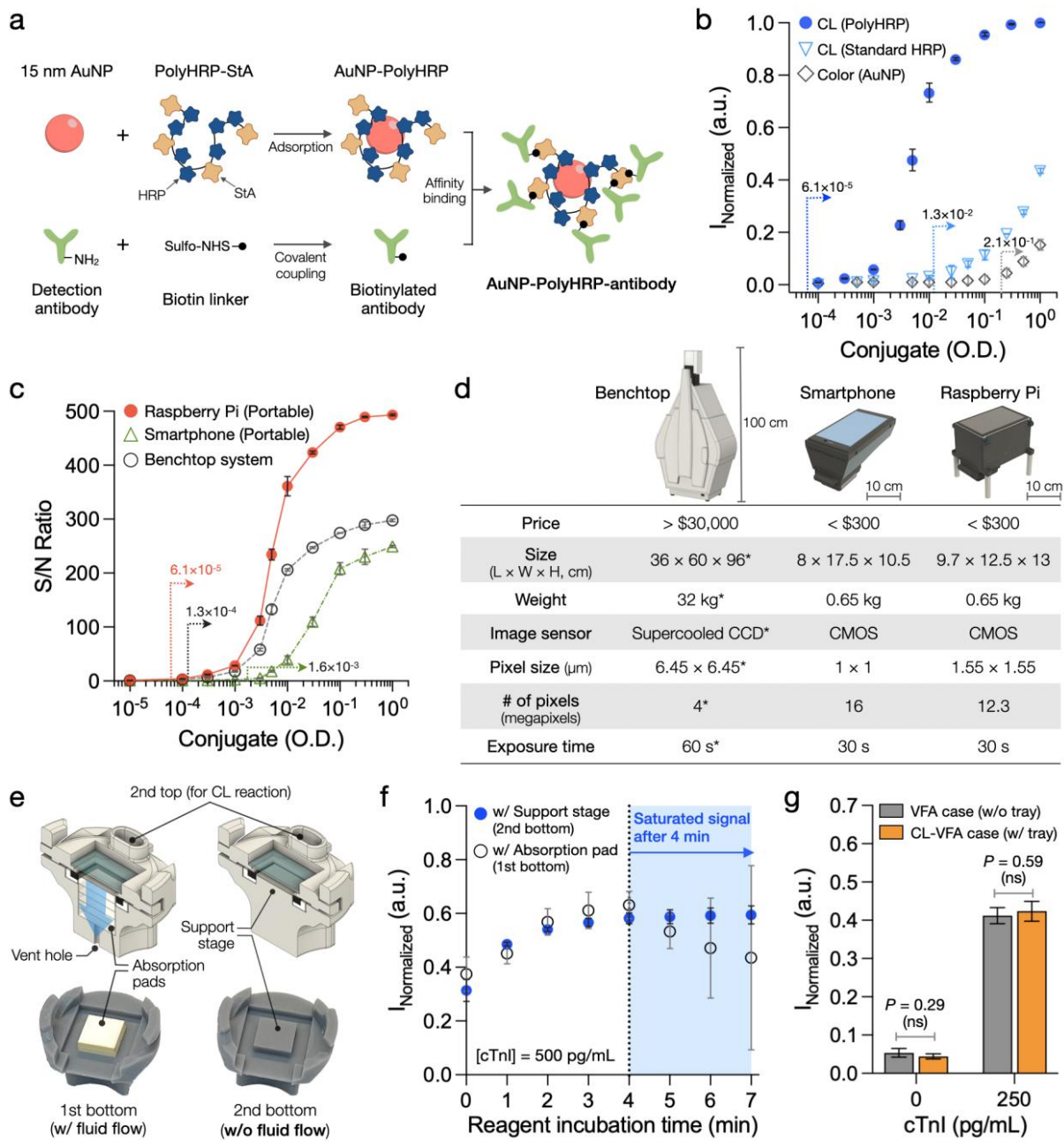

**Figure 2.** Conjugate, reader, and assay cartridge of the CL-VFA. (a) Overview of the conjugation process involving 15 nm AuNP, polyHRP, and biotinylated antibody. (b) Comparison of sensitivity between CL signals generated by AuNP-polyHRP-Ab and AuNP-standard HRP-Ab conjugates, as well as the colorimetric signal of AuNP-polyHRP-Ab conjugate. The conjugates were serially diluted and pre-applied onto the sensing membrane. (c) Signal-to-noise (S/N) ratio comparison of CL signals across different reader systems: the Raspberry Pi-based portable system, a smartphone-based reader, and a traditional benchtop system. (d) Comparative analysis of the key specifications of benchtop, smartphone, and Raspberry Pi-based systems. *The specifications of the benchtop imaging system refer to Chemidoc MP (Bio-Rad), which was used for the comparative study. (e) Design comparison, highlighting the differences between 1st bottom and 2nd bottom cases, with the latter designed to minimize fluid flow during the CL imaging step for improved signal stability. (f) Influence of the reagent flow within the bottom case on CL imaging stability and signal saturation. (g) Comparison of CL assay results using the original VFA case without the tray and the modified tray-based case, demonstrating identical assay performance despite the design modification. The numbers in (b) and (c) are detection cut-offs. Data points in (b), (c), and (g) represent the mean of triplicates ± SD. Data points in (f) represent the mean of duplicates ± SD.

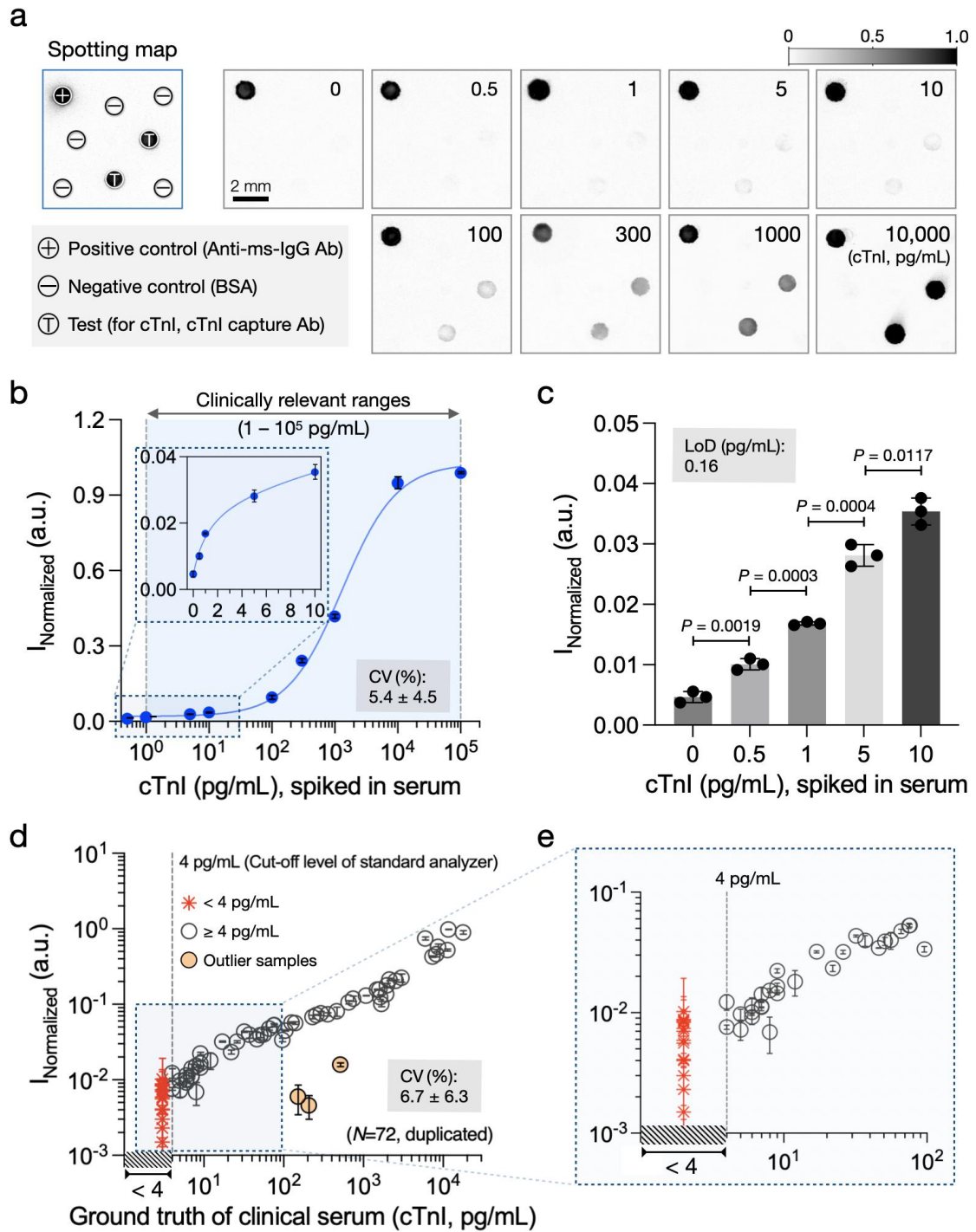

**Figure 3.** Detection performance of the CL-VFA for cTnI spiked serum samples and clinical serum samples. (a) Spotting map and representative sensing membrane images after assaying various concentrations of cTnI and capturing activated assays by the Raspberry Pi-based portable reader. (b) Calibration plot (Blue line; $R^2$ = 0.9929, $y = 0.0147x^{0.4043}$), demonstrating the assay's strong correlation and sensitivity across the clinically relevant range. (c) Comparison of test spot intensities for cTnI concentrations within clinically relevant cut-off levels, highlighting statistically significant differences at various cTnI concentrations. (d) cTnI clinical sample test results. (e) Expanded view of the clinical sample test results below the 100 pg/mL range. Data points in (b) and (c) represent the mean of triplicates ± SD. Data points in (d) and (e) represent the mean of duplicates ± SD.

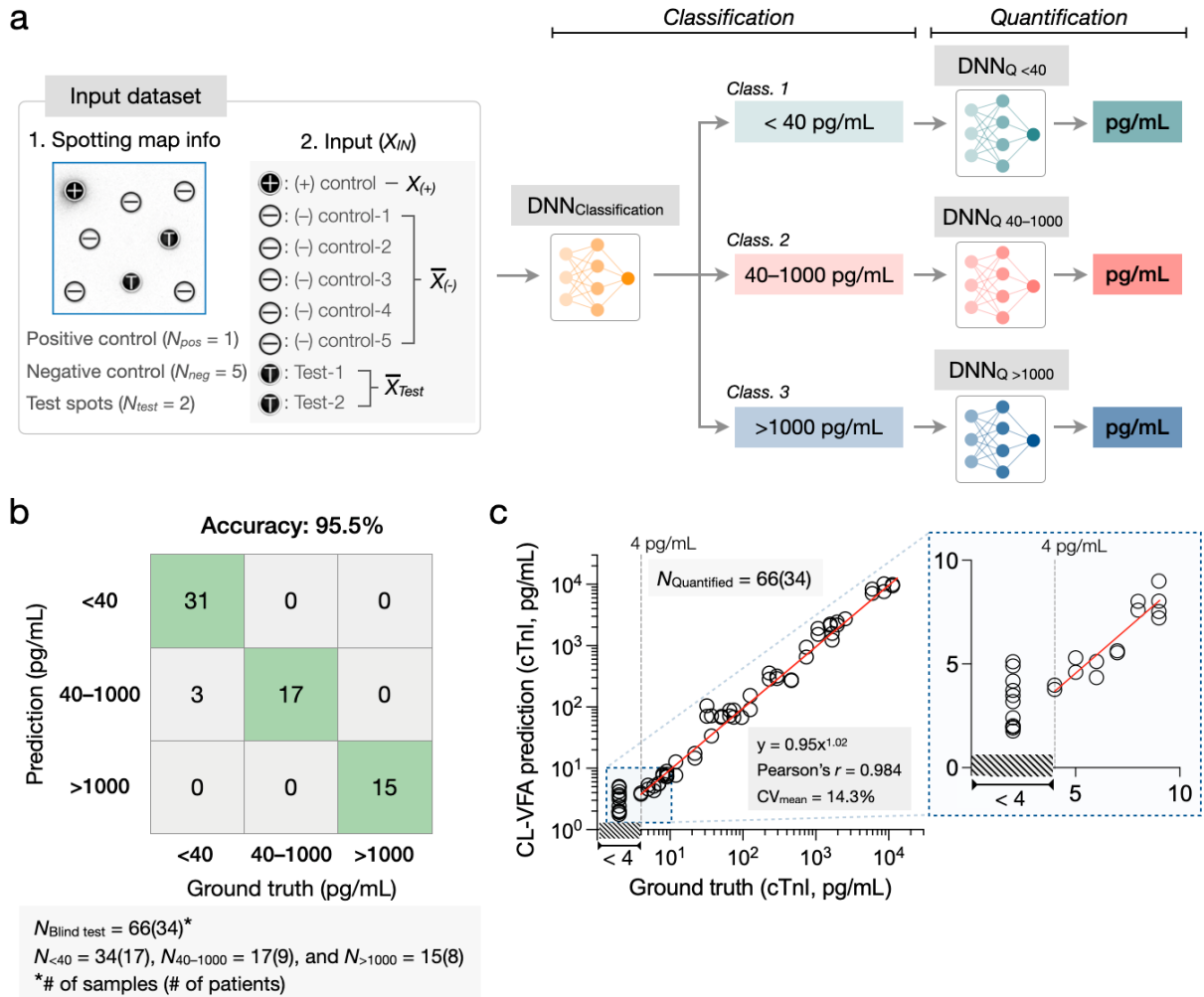

**Figure 4.** Neural network-based analysis of cTnI concentrations in clinical serum samples. (a) Neural network-based cTnI quantification pipeline. The neural network pipeline consists of a classification network ($DNN_{Classification}$) and 3 quantification networks ($DNN_{Q<40}$, $DNN_{Q40–1000}$ and $DNN_{Q>1000}$). $DNN_{Classification}$ classifies all samples into one of the three pre-determined concentration ranges (<40 pg/mL, 40–1000 pg/mL, and >1000 pg/mL) based on the cTnI concentration in the serum sample. Then 3 separate quantification network models ($DNN_{Q<40}$, $DNN_{Q40–1000}$, and $DNN_{Q>1000}$) quantify cTnI concentration in samples from each range. (b) Classification results of $DNN_{Classification}$ for 66 serum samples from 34 patients used in the blind testing set. (c) Combined quantification results from the three quantification models for the 66 serum samples from 34 patients used in the blind testing set. Samples with cTnI concentration in <4 pg/mL range do not have quantitative ground truth labels due to the limitations of the clinically used measurement system, and therefore x-axis does not have quantitative values in that range. All samples with cTnI concentration in <4 pg/mL range used in the blind testing set are stacked above each other within the <4 pg/mL interval.